\documentclass[fleqn,10pt]{wlscirep}
  \usepackage[utf8]{inputenc}
\usepackage[T1]{fontenc}
\usepackage{multirow, booktabs}
\title{Specialization in Criminal Careers}

\usepackage{color}

\author[1,2]{Georg Heiler}
\author[1,3]{Tuan Pham}
\author[1,3]{Jan Korbel}
\author[1,4]{Johannes Wachs}
\author[1,3,5,*]{Stefan Thurner}
\affil[1]{Complexity Science Hub Vienna}
\affil[2]{Technical University Vienna}
\affil[3]{Medical University of Vienna}
\affil[4]{Vienna University of Economics and Business}
\affil[5]{Santa Fe Institute}

\affil[*]{stefan.thurner@meduniwien.ac.at}

\begin{abstract}
We use a comprehensive longitudinal dataset on criminal acts over five years in a European country to study specialization in criminal careers. We cluster crime categories by their relative co-occurrence within criminal careers, deriving a natural, data-based taxonomy of criminal specialization. Defining specialists as active criminals who stay within one category of offending behavior, we study their socio-demographic attributes, geographic range, and positions in their collaboration networks, relative to their generalist counterparts. In comparison to generalists, specialists tend to be older, more likely to be female, operate within a smaller geographic range, and collaborate in smaller, more tightly-knit local networks. We observe that specialists are more intensely embedded in criminal networks and find evidence that specialization indeed reflects division of labor and organization.
\end{abstract}
\begin{document}

\flushbottom
\maketitle

\thispagestyle{empty}

\section*{Introduction}
Criminal careers can be classified in many ways. One useful categorization of criminal behavior is into specialists, describing  individuals who tend to commit the same types of crime, and generalists, who are more versatile in their actions \cite{blumstein1986criminal}. The distinction between specialists and generalists mirrors two major ways criminologists conceptualize the root causes of criminal behavior \cite{delisi2019past}. If criminal behavior is the result of social context \cite{crick1994review}, opportunity \cite{cloward2013delinquency,zembroski2011sociological}, or social learning \cite{burgess1966differential,sutherland1992principles}, we would expect a large degree of specialization in criminal careers. If criminal behavior rather emerges as a result of inherent traits of an individual, for example the capability for restraint and self-control \cite{gottfredson1990general} or the tendency towards psychopathy \cite{delisi2016psychopathy}, one would expect it to manifest as generalist behavior. In practice, both behavioral patterns are observed \cite{blumstein1986criminal}.

This categorization and its conceptual framing are not merely of academic interest. Whether an individual is a (potential) generalist or specialist has implications for their rehabilitation \cite{guerette2005understanding}, especially for juveniles \cite{farrington1988specialization,kempf2017offense}, and for anticipating future criminal behavior. Specialization also seems to play an important role in organized crime \cite{calderoni2012structure,tumminello2021anagraphical,musciotto2022effective} and in gangs \cite{rostami2015myth}, where it facilitates an efficient division of labor. Authorities that seek to interdict criminal networks often target members with high human capital that is difficult to replace \cite{duijn2014relative}. This perspective describes criminal markets in terms of a human value chain, in which goods and services are exchanged \cite{gottschalk2009value}. As in traditional economic value chains, certain stages require specific skills and inputs that are carried out by specialists. To the extent that high human capital specialists are not easily replaced, they represent prime targets for intervention by law enforcement \cite{sparrow1991application}.   

Despite this apparent interest, large-scale empirical evidence on specialization in criminal career trajectories and its correlates are limited, likely owing to the difficulty of procuring and handling relevant data. In this work we present such an analysis based on data covering five years of all criminal \textit{charges} against individuals in a small central European country. Specifically our data includes over $1.2$ million distinct criminal events and nearly $600$ thousand individuals charged. Our work extends an emerging field that applies the methods of data and network science \cite{rostami2015complexity,krebs2002mapping,gerbrands2022effect} to study criminal behavior \cite{luna2020corruption,kertesz2021complexity,wachs2021corruption,campedelli2022machine}. In particular, we report novel results on the nature of criminal specialization using comprehensive databases with new methods and harmonized data from an entire country \cite{tumminello2013phenomenology}.

We first outline a statistical method to group criminal actions into coarser types of crimes in an unbiased way. We quantify the likelihood of transitions between these crime types within criminal careers, meaning the likelihood that a criminal active in one type will become active in another type at a later point in time. We use these transitions to visualize the space of criminal activities as a network of crime types. We then define specialists as those individuals who commit crimes only within one one crime type and carry out a statistical analysis of their socio-demographic attributes, in particular their age and gender. Next, novel results about the relationship between specialization and geographic range of criminals' activities are presented. Finally, we present the relative positions of specialists and generalists within the collaboration network of criminals. We find that specialists tend to have smaller networks around them, albeit with more dense connections between their direct collaborators, than generalists.

\section*{Specialization in Criminal Careers}

The primary challenge in defining specialization of criminal behavior is to group different crime types according to a reasonable taxonomy. Assault and battery, for example, are crimes with legal definitions. To the layman, they are clearly related: no one would be surprised if a criminal was convicted of both offenses over the course of a career. Computer hacking and grave-robbing, on the other hand, are intuitively less likely to be carried out by the same person. While legal codes tend to group different kinds of crimes into reasonable categories, such categorizations are not always useful for describing criminal specialization because they reflect artifacts of the historical evolution of the law \cite{klingenstein2014civilizing}. The structure of criminal codes also differs significantly between countries.

We therefore adopt a statistical approach to group crimes that are carried out by the same criminals within individual careers. We compare the observed distributions of co-occurrence of crimes in the overall population against a null model -- that serves as a statistical benchmark that assumes a randomized distribution of criminal activity across criminals, and that determines the relative frequency of crimes. The grouping of statistically significant co-occurrences of different crime categories within careers create a self-generated data-driven typology of crime types. Our approach extends methods developed by Tumminello et al. \cite{tumminello2013phenomenology} by considering not only the number of crimes but also the number of perpetrators committing two types of crimes. 
%

More explicitly, our method to define specialists takes the following steps. First we define a co-occurrence network of crimes, defined by the legal code, within individual criminal careers. In this network two crimes, for instance assault and computer hacking are connected by a weighted edge counting the frequency that any person is charged with those two crimes. This network is quite dense and needs to be filtered, in other words the edges need to be statistically validated. With a statistically validated network of crimes with statistically validated weights we are able to detect which kinds of crimes often co-appear in the same careers across the entire population. Next, we apply a clustering algorithm, grouping the nodes (corresponding to crimes defined by the legal code) into clusters of crime categories. For instance, one such category includes the crimes of fraud, embezzlement, and forgery; another includes rape and sexual harassment. Finally, use this categorization to define criminals as specialists or generalists: a criminal is a specialist if they only commit crimes from a single category across their observed career, otherwise we say they are generalists. We then carry out an analyses of how these two types of criminals differ, i.e. in terms of their socio-demographic features like age and gender, their geographic mobility, or in how they collaborate with other criminals.

\subsection*{Data and Networks}
The dataset contains any criminal charges brought by the police against individuals in the country from 2015-01-01 until 2021-11-09. In total there were 588k perpetrators charged in 1.2 million distinct crimes. These events take place in the focal country, with only a tiny fraction of the events (0.474\%) took place abroad. Due to data privacy reasons and contractual obligations with the project partner data cannot be shared publicly. 

For each event (criminal act) we observe:
\begin{itemize}
    \item location (political region), $r$
    \item time of the act (date), $t$
    \item category (legal paragraph of the act), $c$
    \item demographics (age and gender at time of arrest), $X_p^{\rm age}$, $X_p^{\rm sex}$
    \item unique, anonymized and persistent identifier of the perpetrator, $p$
    \item a unique identifier of the criminal act $M$
\end{itemize}
{From this data we obtain the following data object, $M_{c,p,r}(t)$, from which we derive all the following. Here we use $M$ as binary classifier for a criminal act, it is 1 if it occurred, and 0 of it did not happen. Note that in reality every $M$ has a unique number, that we do not use here. Also note that one criminal act happening in a particular location at a given time may involve several perpetrators, and may be composed of several crime categories. For example think of a robbery of a house that involved 5 criminals who committed the crimes of robbery and murder.} 
In the following we derive two types of network. Both are projections of the natural bipartite structure of our data in which criminals are charged with crimes. 

From the data object we define a simplified  
matrix with only two indices, $M_{cp} := \sum_{t,r}  M_{c,p,r}(t)$ 
as the number of crimes of type $c$ committed by perpetrator $p$, aggregated over time $t$ and regions $r$. 
Thus, the 
matrix $M_{cp}$ is a bipartite, weighted network, where weights are natural numbers, corresponding to the total number of crimes of type $c$ committed by perpetrator $p$.
For the non-zero unweighted adjacency matrix elements we write $A_{cp} := \min\{M_{cp},1\}$.


\subsection*{Crime-crime transition network} 
\label{sec:act-act-projection}
The first projection is on the crime categories and the resulting network consists of links between two crime categories if there is a criminal who are charged with both categories.
The edges are weighted when more criminals are charged with both categories of crimes. 
In particular, starting from the simplified crime-perpetrator network, $ M_{cp}$, we can use a simple projection to obtain a directed crime-to-crime (C-C) network,where  $\mathbf{N}_{cd} = \sum_{p} A_{cp}A_{dp}$ 
corresponds to the number of perpetrators who committed both crimes $c$ and $d$.

We next establish a statistically validated network by filtering the directed and weighted crime co-occurrence network, based on \emph{hypergeometric filtering} \cite{tuminello11}. We define $\mathbf{N}_{a} = \sum_b \mathbf{N}_{ab}$ as the number of perpetrators who committed crime $a$, and $\mathbf{N} = \sum_a \mathbf{N}_a$ is the total number of perpetrators. For the link between crime categories $a$ and $b$ to be significant, we define the $p$-value as
\begin{equation}
p_{val}(\mathbf{N}_{ab}) = 1 - \sum_{x=0}^{\mathbf{N}_{ab}-1} \frac{{\mathbf{N}_a \choose x} {{\mathbf{N}-\mathbf{N}_a} \choose {\mathbf{N}_b -x}} }{{\mathbf{N} \choose \mathbf{N}_b}} \, , 
\end{equation}
which is the cumulative density function of the hypergeometric distribution at $\textbf{N}_{ab}-1$. It is the probability that out of $\mathbf{N}_a$ and $\mathbf{N}_b$ perpetrators who committed crime of types $a$ or $b$, respectively, there are less than  $\textbf{N}_{ab}$ perpetrators who committed both crimes of type $a$ and $b$.
A link is considered significant at significance level, $p$, if $p_{val}(a,b) < p$. Due to multiple hypothesis testing, we introduce a Bonferroni correction\cite{Bonferroni}, where an adjusted $p$-value, $p/m$, is used, where $m$ is the number of tested hypotheses, which --in our case-- is the number of links of the C-C network. We use this simplest and most conservative approach, also because it was shown that more sophisticated corrections, such as the \v{S}id\'{a}k correction\cite{Sidak}, the Bonferroni-Holm method\cite{Holm}, or the false discovery rate~\cite{FDR}, all yield 
similar community structures \cite{tuminello11,tumminello2013phenomenology}. The so- obtained network is called \emph{statistically validated network} and we denote it by $\mathcal{N}_{ab}$.


This approach does not use the full information contained in the simplified matrix, $M_{cp}$, since it uses the unweighted matrix, $A_{cp}$. Relevant information is lost such as the number of crimes of a given type, say $b$, that were committed by perpetrators committing two types of crimes, say $a$ and $b$. To overcome this, let us define, $\mathcal{P}(a,b)$, as the set of perpetrators who committed crimes of both categories $a$ and $b$. The projection of the bipartite network is therefore defined as  $\mathbf{M}_{ab} = \sum_{p \in \mathcal{P}(a,b)} M_{pb}$, which is number of crimes of type $b$ committed by perpetrators who committed both $a$ and $b$. {Note that here the matrix $\mathbf{M}$ cannot be expressed in terms of matrix multiplication of matrix $M$.}
The resulting C-C network is now directed since in general $\mathbf{M}_{ab} \neq \mathbf{M}_{ba}$.

Again, we exclude links that are not statistically significant. Note that links do not have to be significant in both directions. Therefore we validate if the number of crimes $\mathbf{M}_{ab}$ committed by $\mathbf{N}_{ab}$ perpetrators is statistically significant. To this end, we define $\mathbf{M}_b = \sum_a \mathbf{M}_{ab}$ is the number of crimes of type $b$. We consider a random distribution of $\mathbf{M}_b$ crimes to $\mathbf{N}_b$ perpetrators, where each perpetrator committed at least one crime. Simple combinatorics yields the number of such divisions as ${\mathbf{M}_b-1} \choose{\mathbf{N}_b-1}$. The crimes can be divided into two groups of perpetrators, the ones who also committed crime $a$ and those who did not. Since the number of crimes committed by perpetrators who also committed crime $a$ is $\mathbf{M}_{ab}$, the total probability that $\mathbf{N}_{ab}$ perpetrators out of $\mathbf{N}_b$ who committed  $\mathbf{M}_b$ crimes would commit $\mathbf{M}_{ab}$ crimes can be expressed as
\begin{equation}\label{eq:2}
p(\mathbf{M}_{ab}|\mathbf{N}_{ab},\mathbf{M}_b,\mathbf{N}_b) = \frac{{{\mathbf{M}_{ab}-1} \choose {\mathbf N_{ab}-1}} {{\mathbf{M}_b - \mathbf{M}_{ab}-1} \choose {\mathbf{N}_b - \mathbf{N}_{ab}-1}}}{{{\mathbf{M}_b-1} \choose {\mathbf{N}_b-1}}} \, .
\end{equation}
Thus, the $p$-value corresponding to $\mathbf{M}_{ab}$ is the probability that the number of crimes of type $b$ committed by perpetrators who committed both crimes of type $a$ and $b$ is smaller than $\mathbf{M}_{ab}$, i.e., 
\begin{equation}
p_{val}(\mathbf{M}_{ab}) = 1- \sum_{x= \mathbf{N}_{ab}}^{\mathbf{M}_{ab}-1} \frac{{{x-1} \choose {\mathbf{N}_{ab}-1}} {{\mathbf{M}_b - x-1} \choose {\mathbf{N}_b - \mathbf{N}_{ab}-1}}}{{{\mathbf{M}_b-1} \choose {\mathbf{N}_b-1}}} \, .
\end{equation}
Note that the summation index $x$ goes from $\mathbf{N}_{ab}$ since each perpetrator committed at least one crime of both types, $a$ and $b$. Again, the $p$-value $p$ has to be corrected for multiple hypothesis testing and we use the conservative Bonferroni correction. The statistically validated directed network is denoted as $\mathcal{M}_{ab}$.

For computational purposes, since factorials of large numbers are computationally hard to obtain, it is convenient to express the hypergeometric distribution in Eq.~\eqref{eq:2} to use the recursive formula 
\begin{equation}
p(\mathbf{M}_{ab}+1|\mathbf{N}_{ab},\mathbf{M}_b,\mathbf{N}_b) = \frac{\mathbf{M}_{ab}\, (\mathbf{M}_b-\mathbf{M}_{ab}-\mathbf{N}_b+\mathbf{N}_{ab})}{(\mathbf{M}_{ab}-\mathbf{N}_{ab}+1)\, (\mathbf{M}_b-\mathbf{M}_{ab}-1)}\,  p(\mathbf{M}_{ab}|\mathbf{N}_{ab},\mathbf{M}_b,\mathbf{N}_b)\, .
\end{equation}
The probability of the lowest possible $\mathbf{M}_{ab}=\mathbf{N}_{ab}$, which reads 
\begin{equation}
p(\mathbf{N}_{ab}|\mathbf{N}_{ab},\mathbf{M}_b,\mathbf{N}_b) = \frac{(\mathbf{N}_b-1)!(\mathbf{M}_b-\mathbf{N}_{ab}-1)!}{(\mathbf{M}_b-1)!(\mathbf{N}_b-\mathbf{N}_{ab}-1)!}\, ,
\end{equation}
can be efficiently calculated by the method of decomposition of factorials into prime numbers \cite{rankin_1983}. The procedure is analogous to the one used for efficient calculation of original hypergeometric distributions \cite{Wu93}.

\subsection*{Clustering of crime types}

Given the statistically validated networks we now can use community detection algorithms to classify crimes according to their co-appearance. We extend  existing methodology by involving not only the number of crimes but also the number of perpetrators that commit both types of crime to define the statistically validated directed networks.

\subsubsection*{Community detection}
To detect communities in the C-C network, we use the {\em Infomap} algorithm \cite{rosvall2009map} that is based on random walks on the network. One could equally employ other community detection methods such as the Louvain \cite{louvain} or Leiden method\cite{leiden}. We use the Infomap algorithm, to be able to compare results directly with Tuminello et al.\cite{tumminello2013phenomenology} who used the same method. 

The resulting community structure is displayed as a community-community network, where each node represent one community, $\alpha$. We denote the set of nodes as $\alpha,\beta,\dots$,{ where each community has its members (crime categories), e.g., $\alpha =\{c_1,\dots,c_k\}$.} 
The undirected community-community network is given by
$\mathbf{C}_{\alpha\beta} = \sum_{a \in \alpha, b \in \beta} \mathcal{N}_{ab}$.
Thus the link weights correspond to the total number of perpetrators that committed crimes from both communities. We also define the directed links obtained from the statistically validated directed community-community networks as
$\mathbf{D}_{\alpha\beta} = \sum_{a \in \alpha,b \in \beta} \mathcal{M}_{ab}$.

\subsection*{Criminals' trajectories and identifying level of specialization for communities of crimes}

To define generalists and specialists we 
calculate a crime trajectory for each perpetrator, a sequence of crime types from the community-community network. For each crime committed by a given perpetrator, we assign a crime type obtained from the community detection on the statistically validated crime-crime network. We denote a {\em trajectory} of a perpetrator $p$ as $x^p_t\in \mathcal{C}$, which indicates that $p$ committed crime $x^p_t$ at time $t$. By taking into account all perpetrators who committed more than one crime, we estimate the transition frequencies between the crime communities. To first order, transitions between crime communities can be described as Markov chains with transition probabilities, $p(x_{t+1} \in \alpha|x_{t} \in \beta)$. 
By using the local mutual information\cite{gueguen2014local}, $I_{\alpha \to \beta}$, between communities
\begin{equation}
I(\alpha \rightarrow \beta) = \log_2 \frac{p(x_{t+1} \in \beta|x_t \in \alpha)}{ p(x_{t+1} \in \beta)} =\log_2 \frac{p( x_{t+1} \in \beta,x_t \in \alpha)}{  p(x_{t+1} \in \beta)\cdot p(x_t \in \alpha)}
\equiv \log_2 \frac{p(\alpha \to \beta)}{p(\beta) \cdot p(\alpha)}\end{equation}
we determine how often (compared to a random jump model) perpetrators jump from crime community $\alpha$ to crime community $\beta$. Assuming the distribution being stationary, one can omit the time index and denote the probability of observing crime from cluster $\alpha$ simply as $p(\alpha)$ and observing transition $\alpha \to \beta$ as $p(\alpha \to \beta)$.
For the case, where we observe no jump $\alpha \to \beta$, the mutual information is minus infinity. In this case, we let $I(\alpha \to \beta)$ be undefined.

Calculating $I(\alpha \to \beta)$ allows us to compare the Markov chain model to a null model of random jumps according to the probability of committing crime from community $\beta$ given by $p(x_{t+1} \in \beta)$. This local information is not symmetric in its arguments since the former crime class denotes the source community and the latter class denotes the target community. In general, the local mutual information can be both positive and negative. If $I(\alpha \rightarrow \beta) >0$, the frequency of jumps between the groups is higher than expected from the null model, if $I(\alpha \rightarrow \beta) <0$ the frequency is smaller. Particularly interesting is the local mutual information of transition within one community, i.e., $I(\alpha) := I(\alpha \to \alpha)$, which measures the tendency of a perpetrator to remain in the community when committing two subsequent crimes. The value of $I(\alpha)$ means that $p(\alpha \to \alpha) = \left(2^{\frac{I(\alpha)}{2}} p(\alpha)\right)^2$, so according to the null model of random jumps, the frequency of observing crimes from cluster $\alpha$ is rescaled by a factor $2^{\frac{I(\alpha)}{2}}$, so $I(\alpha)/2$ is the rescaling exponent determining the deviation from the null model. 
Intuitively, the mutual information quantifies, how often the perpetrators commit two consecutive crime categories after each other compared to the situation, where they commit the crimes randomly. 
Particularly, the intra-community mutual information $I(\alpha)$ tells us how much more (or less) often a perpetrator commits two consecutive crimes in one crime cluster compared to the probability that two consecutive crimes committed will be from same crime cluster. 
Intuitively, the mutual information measures how much more (or less) we observe that two consecutive crimes of one perpetrator will be from crime clusters $\alpha$ and $\beta$, compared to the situation when the two crimes are committed by two independent perpetrators. Particularly, the value $I(\alpha)$ tells us that the one particular perpetrator commits two consecutive crimes from cluster $\alpha$ at least $2^{I(\alpha)/2}$ more often than two distinct randomly chosen perpetrators. 


\subsection*{Definition of Specialists and Generalists}
We are now ready to use the obtained clustering of crime types to our primary goal of identifying specialists and generalists. We define a specialist as any criminal who has been charged with crimes from only 1 of the 21 identified clusters. Criminals charged with crimes from multiple clusters within their career are considered generalists. In subsequent analyses of the differences  between specialists and generalists, we shall focus on a subset of the data consisting of {\em repeat-offenders}. This consists of a subset of 64,406 individuals that were charged with at least five crimes in the dataset spanning five years. 

\subsection*{Geographic Range of Criminal Activity}

Although most criminals tend to operate within a limited geographical range, some seem to be more effective when diversifying the locations of their activity \cite{rhodes2017crime}. We define the geographic range of an individual by calculating a quantity known as the {\em radius of gyration} on the locations of their criminal activities. Individuals committing crimes consistently in the same locations have a low, more mobile ones --with activities in various regions-- have a high radius of gyration. 
 
For each offense we geolocate the region's latitude/longitude centroid, $r$. We generate a vector of offense locations for each perpetrator $i$: $\vec{r}_{i\mu} = (x_{i\mu}, y_{i\mu})$, at location index $\mu = 1 ... N_{locations}$, where $x$ and $y$ represent longitude and latitude, respectively. We calculate the average location as where the perpetrator is usually active, comparing the individual crime locations $\vec{r}_{i\mu}$ to the centroid of the criminal's history $\overline{r}_i = \frac{\sum_\mu \vec{r}_{i\mu} }{\sum_\mu }$.

The \emph{radius of gyration} $R_G$ is calculated as the square root of the mean of the squared distances, $d$, (calculated as the Haversine distance, which calculates a distance in meters from latitude and longitude coordinates given in degrees) of the locations $\vec{r}_{i\mu}$ to the individual's centroid $\overline{r}_i$:
\begin{equation}
	R_{\mathrm{G},i} = \sqrt{\frac{\sum_\mu d(\overline{r}_i, \vec{r}_{i})^2}{\sum_\mu}} \, . 
\end{equation}

\subsection*{Collaboration network.}

We derive a second network from the dataset that maps the collaboration between criminals. Specifically, nodes in this network are individual criminals, who are connected by an edge if they collaborated on a specific crime in our database. 
Edges represent those criminals that were charged for the same criminal act. Since individuals can be charged for the same criminal act that they committed together, for example by being arrested and charged for the same burglary event, we can construct a network of collaborations where nodes are criminals and edges indicate if they were charged for the same act, with a weight on the edge increasing in the frequency of collaboration.
We define the collaboration network as a matrix ${\cal C}$. The entry ${\cal C}_{cd}$ quantifies the collaboration between criminals $c$ and $d$. Specifically, ${\cal C}_{cd}=\sum_k \dfrac{\delta^{k}_{c}\delta^{k}_{d}}{n_{k}-1}$, where $\delta^{k}_{c}$ is equal to 1 if criminal $c$ participated in crime event $k$, $n_{k}$ is the number of criminals involved in crime event $k$, and the sum is over all crime events in the dataset.

This edge weighting method is sometimes called Newman's hyperbolic weighting method~\cite{newman2001scientific}, in which the contribution of a specific criminal collaboration to the weight between two criminals is inversely proportional to the number of collaborators on that crime. For example if two criminals collaborate on a specific crime alone as a pair, the edge between them will have a higher weight than that between two criminals who collaborate on a specific crime with ten other collaborators. We construct the collaboration network including all individuals charged with a crime in our dataset. In the analysis below we contrast the characteristic positions and connectivity patterns of specialists and generalists in this collaboration network.

\section*{Results}

\subsection*{Crime clusters}


The 21 identified crime clusters, i.e., the nodes in the community-community network $\mathbf{C}_{\alpha \beta}$, from the clustering of the statistically validated C-C network, $\mathcal{N}_{ab}$, are summarized in Tab. \ref{tab:1}. Since crimes naturally cluster according to crime domains, we label the clusters by designations such as, "economic crimes", "violent crimes", "street criminality", etc. The table contains the number of crime types (paragraphs in the criminal code) that belong to the community, number of delicts committed there, and the number of involved perpetrators.  Several example crimes that belong to the community are mentioned in the rightmost column.  

\begin{figure}
    \centering
      \includegraphics[width=\linewidth]{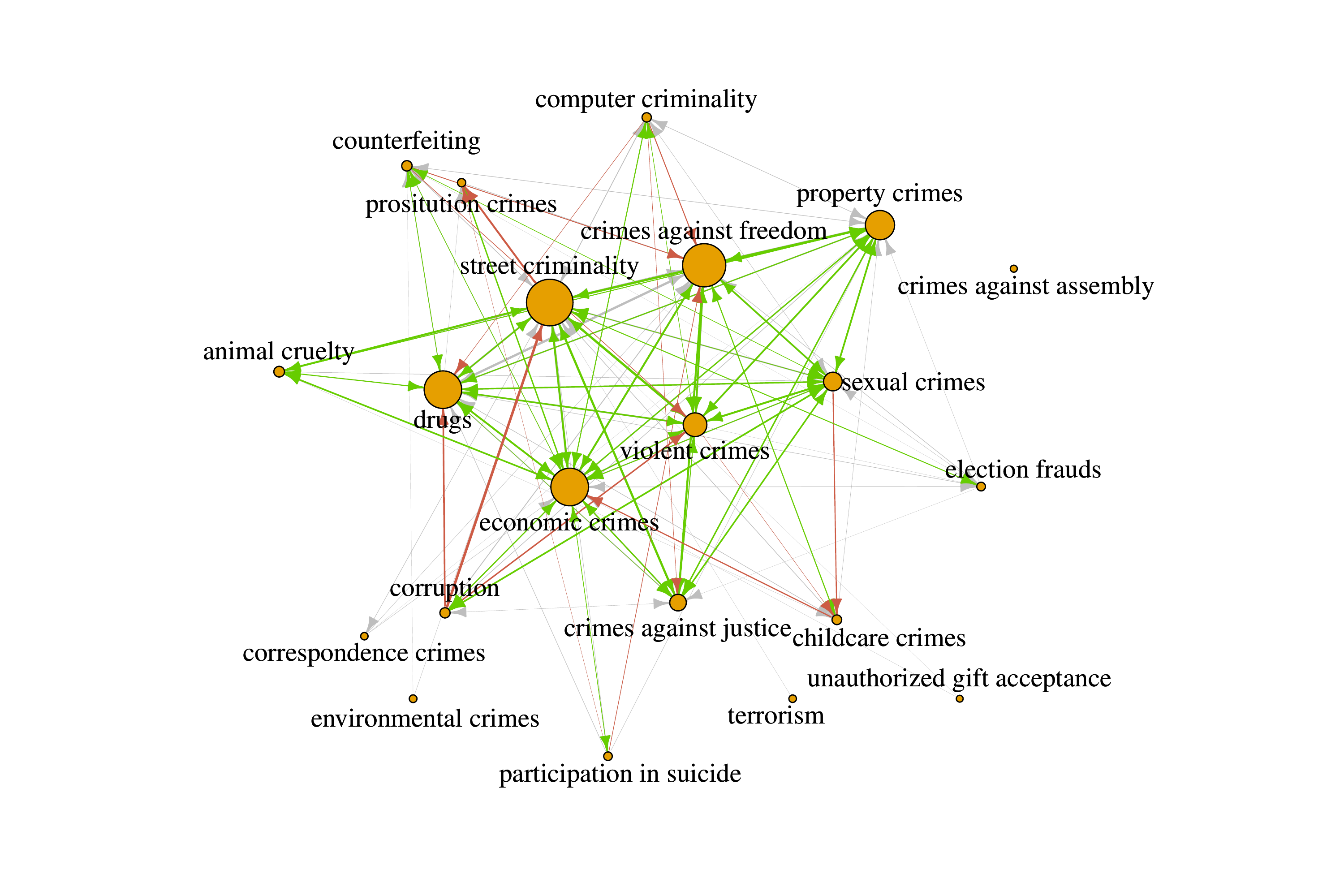}
    \caption{\textbf{Transitions between crime clusters.} 
    Nodes represent the crime clusters, their size corresponds to the number of delicts. Green arrows indicate that all three link types, i.e., links between communities calculated from the underected community-community network
    $\mathbf{C}_{\alpha\beta}=\mathbf{C}_{\beta\alpha}$, and directed links from the directed community-community network in both direction, i.e.,  $\mathbf{D}_{\alpha\beta}$, and $\mathbf{D}_{\beta\alpha}$, are all statistically significant. The link width represents the number of delicts committed by the perpetrators in both crime clusters, $\alpha$ and $\beta$. Green arrows are typically observed between large crime clusters as e.g., {"street criminality"}, {"crimes against freedom"}, and {"violent crimes"}.
    Red arrows represent statistically significant links in the undirected network, $\mathbf{C}_{\alpha\beta}=\mathbf{C}_{\beta\alpha}$, and a link of the directed (validated) network in one direction only (i.e., $\mathbf{D}_{\alpha\beta}$ is significant but $\mathbf{D}_{\beta\alpha}$ is not). These are typically observed between from large to small crime clusters, as e.g., {"corruption"} to {"street criminality"}, {"prostitution"} to \emph{"crimes against freedom"}, or {"sexual crimes"} to {"childcare crimes"}.
    Gray arrows represent links where a link in the directed community-community network $\mathbf{D}_{\alpha\beta}$ is significant but the corresponding link in the undirected community-community network $\mathbf{C}_{\alpha \beta}$ is not. 
    }
    \label{fig:2}
\end{figure}

The crime community-community network is depicted in Fig. \ref{fig:2}. Nodes are the crime clusters, the size of the nodes correspond to the number of crimes committed, and the link width represents the amount of crimes committed by perpetrators that committed crimes in both clusters. The connection between a pair of crime clusters, say $\alpha$ and $\beta$, can be described by three numbers: the value of the undirected community-community network link $\mathbf{C}_{\alpha\beta} = \mathbf{C}_{\beta \alpha}$, and the two directed links of the directed community-community network, i.e., $\mathbf{D}_{\alpha\beta}$ and $\mathbf{D}_{\beta \alpha}$. 
Depending on whether the links are significant or not (with respect to the hypergeometric filtering), we can divide the links between communities into three categories.

First, the links where all $\mathbf{C}_{\alpha\beta}$, $\mathbf{D}_{\alpha\beta}$ and $\mathbf{D}_{\beta\alpha}$ are statistically significant. These links show the strong relations between the clusters in both directions and are depicted in green in Fig. \ref{fig:2}. A typical example of such a link is "street criminality" and "crimes against freedom". The second type are those links where $\mathbf{C}_{\alpha\beta}$ and $\mathbf{D}_{\alpha\beta}$ are significant, but $\mathbf{D}_{\beta\alpha}$ is not significant. These are most interesting since they enable us to reveal the structure of the link, as the link indicates that the perpetrators committing crimes from cluster $\alpha$ also commit crimes from cluster $\beta$ but not vice versa. These links are depicted in red in Fig.  \ref{fig:2}. Finally, the third case is when either $\mathbf{D}_{\alpha\beta}$ or $\mathbf{D}_{\beta\alpha}$ are significant but $\mathbf{C}_{\alpha\beta}$ is not significant. These links are depicted in gray and we do not consider them as relevant. 

We observe six large crime clusters, i.e., "economic crimes", "crimes against freedom", "street criminality", "drug crimes", "violent crimes", and "property crimes". These clusters appear in the center of the community network; the strongest connections are between "street crimes", "violent crimes" and "crimes against freedom", constituting a strong triangle. "Crimes against freedom" have a strong link to "property crimes". Similarly, there are strong links between "street criminality", "drugs" and "economic crimes". All links are bi-directional and these six clusters are strongly connected. The remaining clusters are connected only to a few other clusters; some of the links are uni-directional and typically go from smaller to larger communities. We mention a few examples: (a) a link from "computer criminality" to "drugs", which might point to online drug sales, (b) links from "sexual crimes" and "violent crimes" to "childcare crimes", which corresponds to child sexual abuse, and "violence against children", respectively, (c) links from "corruption" to "drugs" and "street criminality", pointing to the fact that corruption is often connected with other criminality where criminals try to bribe police officers or witnesses, (d) a link from "street criminality" to "prostitution" and, consequently, a link from "prostitution" to "crimes against freedom" and "participation in suicide". 

\begin{table}[t]
    \centering
    \begin{tabular}{|l||r|r|r |l|}
\hline
Community name  & \# crime types & \# delicts & \# perpetrators & Example crimes \\
      \hline
      Economic crimes  &  54 & 238,221 & 160,772 & fraud, embezzlement, forgery \\
      Crimes against freedom & 15 & 331,138 & 254,165 &  coercion, threat, battery\\
      Street criminality & 17 & 394,491 & 199,929 & theft, suppression of documents\\
      Drug crimes & 15 & 236,980 & 150,300 & drug possession and trafficking\\
      Violent crimes & 20 & 73,241 & 63,771 & murder, robbery, mayhem\\
      Sexual crimes & 24 & 36,701 & 31,687 & rape, sexual harassment\\ 
      Property crimes & 13 & 127,340 & 84,318 & property damage, arson\\
      Crimes against justice & 11 & 25,372 & 23,531 & perjury, insurance fraud\\
      Corruption & 12 & 3,612 & 3,229 & bribery, abuse of power\\
      Computer criminality & 11 & 2,170 & 2,008 & misuse of data (private/banking/corporate)\\
      Prostitution crimes & 6 & 923 & 836 & pimping\\
      Animal cruelty & 5 & 4,592 & 4,175 & animal cruelty\\
      Election frauds & 6 & 1,321 & 1,094 & election falsification\\
      Childcare crimes & 5 & 2,732 & 2,603 & child neglect, child abduction\\
      Terrorism & 4 & 343 & 324 & terrorism financing, terrorism approval\\
      Counterfeiting & 3 & 4,005 & 3,080 & counterfeiting\\
      Participation in suicide & 4 & 1,194 & 1,185 & facilitating suicide\\
      Environmental crimes & 3 & 431 & 378 & intentional damage to the environment\\
      Crimes against assembly & 2 & 95 & 95 & preventing/disturbing an assembly\\
      Unauthorized gift acceptance & 2 & 47 & 45 & acceptance of gifts by authorities\\
      Correspondence  crimes & 1 & 261 & 239 & violation of the secrecy of correspondence \\
      \hline
      Total & 233 & 1,485,641 & 987,764 &  unique perpetrators: 581,486\\
      \hline
    \end{tabular}
    \caption{\textbf{Characterization of crime clusters of the statistically validated crime network.} Each community is characterized by its crime composition, number of crime types (corresponding to the specific paragraph of the crime code), number of delicts, and number of perpetrators. The rightmost column provides several examples of representative crimes within every crime cluster. Note that the number of unique perpetrators is not the sum of the perpetrators in each cluster since some of the perpetrators were committing crimes in several crime clusters.}
    \label{tab:1}
\end{table}

\subsection*{Criminal trajectories and level of specialization of criminal communities}

\begin{figure}[t]
    \includegraphics[width=\linewidth]{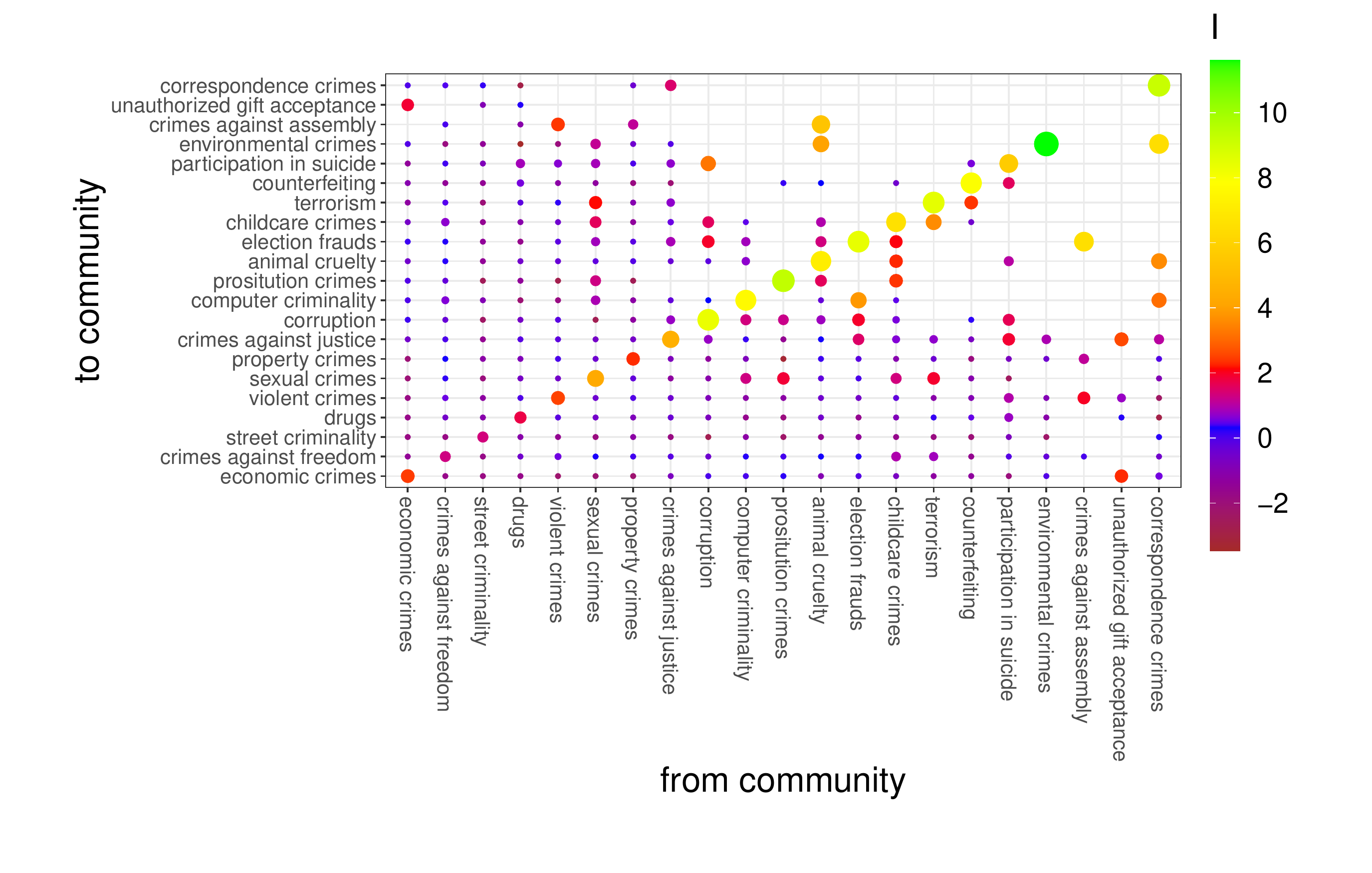}
    \caption{\textbf{Local mutual information $I(\alpha \to \beta)$ between crime communities.} {It indicates how much more (less) often a perpetrator commits two consecutive crimes from crime clusters $\alpha$ and $\beta$, compared to the frequency of committing crimes from crime clusters $\alpha$ and $\beta$.} The local mutual information is encoded both by color (see the color-scale) and by size (for positive $I$, the size of the point is proportional to $I$.)
    High (orange to green) values on the diagonal highlight those crime types which criminals tend to stay within, suggesting specialization.}
    \label{fig:3}
\end{figure}

We calculate the local mutual information, $I(\alpha \rightarrow \beta)$, for the 21 crime clusters. Here we use a reduced dataset, where from the total of $581,486$ different perpetrators in the data, we look at the subset of $131,409$ who committed more than one crime. Results are depicted in Fig. \ref{fig:3}. The local mutual information is encoded both by color (see color-scale) and size (for positive $I$, the larger the local mutual information, the larger the dot). A special role is played by the diagonal of the matrix, i.e.,  $I(\alpha \rightarrow \alpha) = I(\alpha)$ that represents the relative frequency of continuing in the criminality of the same type. Individuals who stay within the same type were identified as \emph{specialists}. We observe that $I(\alpha) >0$ for all communities, which means that committing two subsequent crimes in the same  cluster is more probable than in the null model, which is in agreement with the crime clusters from the previous section.

The value of $I(\alpha)$ changes considerably between different communities. This allows us to associate crime clusters with {several types}. Remember that $I(\alpha)/2$ represents the characteristic rescaling exponent of a crime cluster,  $\alpha$. To obtain a threshold for the distinction between crime clusters of crimes committed by generalists and specialists, we choose $I_{crit}= 4$, so $2^{(I_{crit}/2)} =4$, which means that for specialists we have $p(\alpha \to \alpha) \geq (4 p(\alpha))^2$. { So the probability that a perpetrator commits two consecutive crimes from one crime cluster $\alpha$ is at least 16 times as as high as the probability that 
the two crimes are committed by two random perpetrators.
}

We obtain that crime clusters with $I(\alpha)<I_{crit}$ are: {"economic crimes"} ($I=2.42)$, {"crimes against freedom"} ($I=1.27$), {"street criminality"} ($I=1.31$), {"drug-related crimes"} ($I=1.73$), {"violent crimes"} ($I=2.50$), and {"property crimes"} ($I=2.29$).   These crimes are therefore typically committed by generalists.

On the other hand, crimes with $I(\alpha) > I_ {crit}$ include {"sexual crimes"} ($I=4.27$), {"crimes against justice"} ($I=4.49)$, {"corruption"} ($I=8.32$), {"computer criminality"} ($I=7.60$), {prostitution} ($I=9.30$), {"animal cruelty"} ($I=7.13$), {"election frauds"} ($I=8.39$), {"childcare crimes"} ($I=6.57$), {"terrorism"} ($I=8.47$), {"counterfeiting"} ($I=7.91$), {"participation in suicide"} ($I=5.66$), and {"environmental crimes"} ($I=11.59$). Finally, the remaining crime communities, i.e., {"crimes against assembly"}, {"unauthorized gift acceptance"}, and {"correspondence crimes"}, are so rare that the number of transitions is too small to make a valid classification.

Further, we observe that in several cases, the local information is significantly positive between different clusters, i.e.,  $I(\alpha \to \beta) \geq 2$, where $\alpha \neq \beta$. For example, we observe significant local levels of mutual information between {"economic crimes"} and {"unauthorized gift acceptance"}. This link was observed also in Fig. \ref{fig:2}, but it was not considered  significant in the network $\mathbf{C}_{\alpha\beta}$, probably due to the fact that the latter cluster is observed quite rarely (only 47 delicts, see Tab. \ref{tab:1}). Moreover, we observe that {"prostitution crimes"} and {"sexual crimes"} have also a significant local mutual information, which is plausible due to the common sexual nature of both clusters. Most interesting are the cases when $I(\alpha \to \beta)$ is much higher than $I(\beta \to \alpha)$. To these  significantly asymmetric transitions belong: {"election frauds"} $\to$ {"computer criminality"}, {"animal cruelty"} $\to$ {"environmental crimes"} and {"crimes against assembly"}, {"crimes against assembly"} $\to$ {"election frauds"}, and {"corruption"} $\to$ {"participation in suicide"}. An interesting aspect here is that some of the links are not significant when compared with the links between crime clusters shown in Fig. \ref{fig:2}. 


\subsection*{Characteristics of Specialists vs. Generalists}

Given the crime clusters we can classify individual perpetrators as generalists or specialists according to our definition: specialists are those individuals staying within a single crime cluster across their career. Among the $64,406$ repeat offenders (defined above as those individuals charged five times in our data), we categorize $11,212$ (17\%) individuals as specialists and $53,194$ (83\%) as generalists.

\subsubsection*{Socio-demographic differences of Specialists vs Generalists }

We can now assign other socio-demographic information to criminals and provide statistical evidence of over- and under representation of specific traits in the two respective populations. We describe differences in observed behavior between the two groups, for example the geographic diversity of their criminal actions, and their collaboration activity. We observe significant differences in specialization behavior based on gender and age. Women are significantly more likely to be specialized than men (26\% of women vs 16\% of men, $p<.01$ Mann-Whitney U). Individuals under the age of 20 are highly versatile, with a specialist rate of only 11\%, vs 13\% for those between the age of 20 and 30. 21\% of individuals older than 30 are specialized. These findings confirm previous empirical findings from the literature carried out at smaller scales, which we review in the Discussion.



\subsubsection*{Mobility of Specialists vs Generalists }

\begin{figure}
    \centering
    \includegraphics[width=0.7\linewidth]{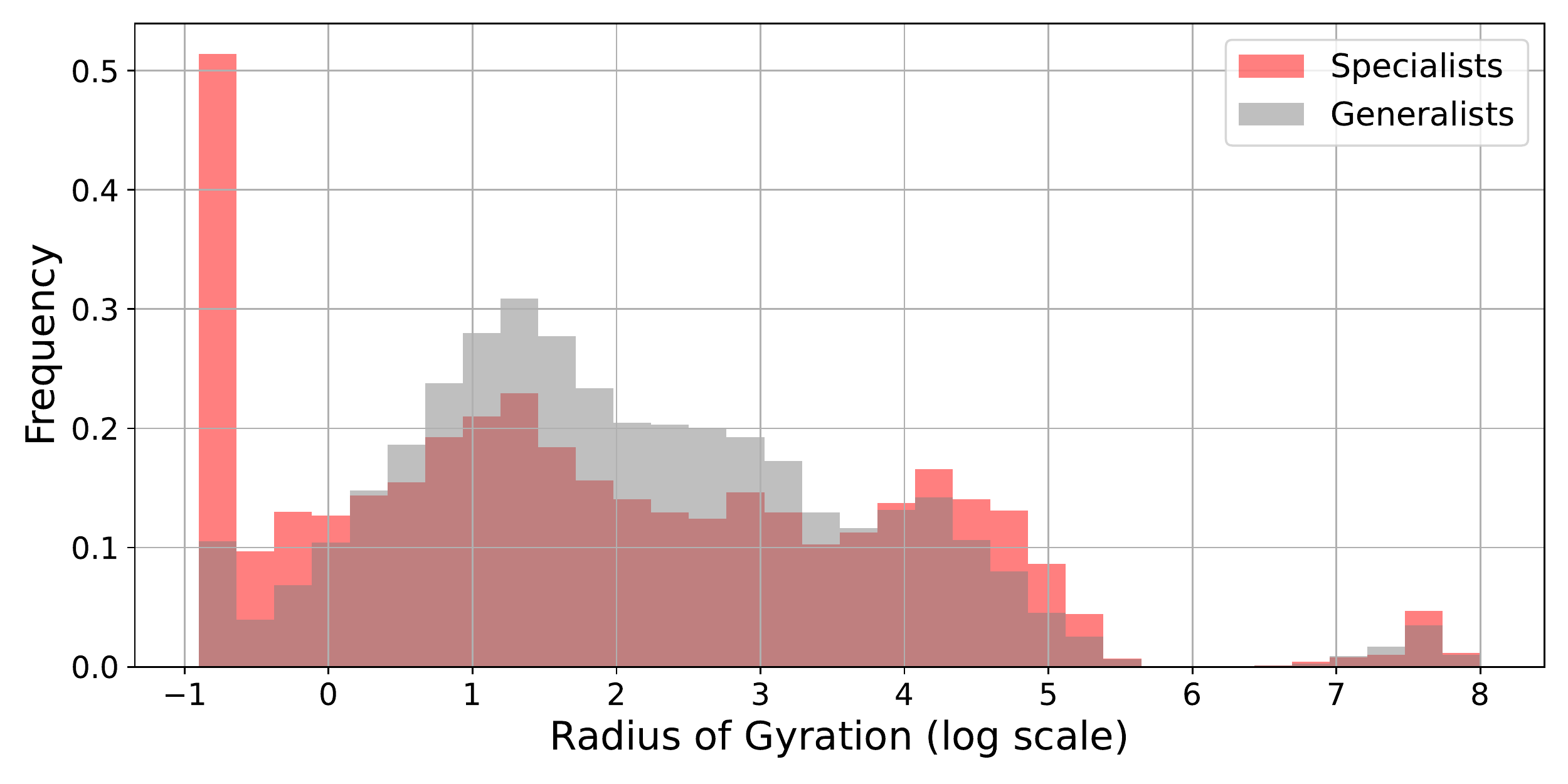}
    \caption{\textbf{Distribution of the radius of gyration of repeat-offender specialists (red) and generalists, on a logarithmic scale. We see that the majority of specialists are charged with crimes in the same place and --if they move-- they tend to move slightly larger distances. Generalists, on average, show greater mobility.} }
    \label{fig:rogtw}
\end{figure}

Specialization has a significant influence on the geographic range of action of  individual perpetrators. Figure~\ref{fig:rogtw} shows the distribution of the (log) radius of gyration, $R$, for specialist (red) and generalist (gray) repeat-offenders. We observe that specialists are far more likely to commit crimes in the same place and that generalists tend to be more mobile. 


To exclude the possibility that 
the observed difference is a statistical artifact from the situation that if generalists are more active (i.e. commit more crimes), they may --by chance-- commit crimes in a greater variety of locations we test for the statistical significance of this difference, while controlling for overall activity. We fit a linear regression model predicting a criminal $i$'s (log) radius of gyration of the form
\begin{equation}
\log R_{\mathrm{G},i} = \beta_{0} + \beta_{1} S_{i} + \beta_{2} \log N_{i} + \epsilon  \, ,
\label{eq:rogtw}
\end{equation}
where $S_{i}$ is a binary variable that is $1$ if the criminal $i$ is a specialist and $0$ if he is a generalists. Here $N_{i}$ is the number of crimes in the observed career of criminal, $i$, $\beta_{0}$ is an intercept, and $\epsilon$ is the error term. Results are in Table \ref{tab:rog_reg}. Specialists tend to operate in a much more geographically confined area. Controlling for how many crimes they commit, specialist radius of gyration is on average 18\% lower than that of generalists.

\begin{table}
\begin{tabular}{lcccccc}
\toprule
                         & \textbf{Coefficient} & \textbf{std. err.} & \textbf{t-statistic} & \textbf{P-value} &  \\
\midrule
\textbf{Intercept ($\beta_0$)}       &       1.15  &        0.025     &    45.7  &         $<0.001$         \\
\textbf{Specialist ($\beta_1$)}      &      -0.19  &        0.017     &   -11.3  &       $<0.001$           \\
\textbf{log(\# Crimes) ($\beta_2$)} &       0.45  &        0.011     &    40.5  &       $<0.001$          \\
\midrule
\textbf{Observations} & 64406 & &  \textbf{Adj. $R^{2}$} & 0.027  \\ 
\bottomrule
\end{tabular}
\caption{Linear regression (OLS) results predicting individuals' criminal log radius of gyration. We report the estimated coefficients of equation \ref{eq:rogtw}, their standard errors, $t$-statistic and the resulting $p$-values. Controlling for the number of crimes, specialists have a radius of gyration of around 18\% lower than generalists on average.}
\label{tab:rog_reg}
\end{table}

\subsection*{Position of Specialists and Generalists in the Collaboration Network}

These differences in characteristics of specialists and generalists led us to ask whether there are significant differences in how they actually cooperate with other criminals. After all, if criminal specialization is the result of learning and socialization, rather than an innate tendency of individuals, there are likely to be significant differences in how specialists and generalists interact. The relative positions of specialists and generalists in the network of criminal collaboration can suggest how human capital (i.e. specialization) and social capital (i.e. important network positions) coordinates in criminality \cite{morselli2008brokerage,duijn2014relative}.

We report a summary statistics about the positions of specialists and generalists in the collaboration network, ${\mathcal {C}}$,  described in the methods in Tab. \ref{tab:summstat}. For each criminal with at least five offenses in the last five years, we derive the following network characteristics:
\begin{itemize}
    \item \textit{Degree}: number of collaborators.
    \item \textit{Strength}: number of collaborations, counting repeated collaborations with alters.
    \item \textit{2-step neighbors}: number of criminals within two steps of a criminal.
    \item \textit{clustering coefficient}: share of pairs of neighbors of the criminal that are connected themselves 
    \item \textit{Has-network-connection}: if criminal has any connection at all.
    \item \textit{Strength/degree}: ratio of strength to degree of the criminal.
\end{itemize}
We find that generalists tend to have larger ego networks than specialists: more direct connections (degree) (mean 2.98 vs 2.11, Mann-Whitney U $p$-value $<0.01$) and 2-step neighbors (mean 5.27 vs 11.74, Mann-Whitney U $p$-value $<0.01$), both on average and the median level. Generalists are slightly more likely to have  collaborations than specialists (58\% vs 66\% - Mann-Whitney U $p$-value $<0.01$). Specialists have more repeated connections (average strength 5.49  vs 3.26; a higher strength to degree 1.57 vs 0.66; both have a significant Mann-Whitney U $p$-value of  $<0.01$). Specialists tend to have slightly more closed networks, as seen in the average clustering coefficient of 0.27 vs 0.23 for generalists (significant difference at $p<0.01$). 

\begin{figure}
    \centering
    \includegraphics[width=0.9\linewidth]{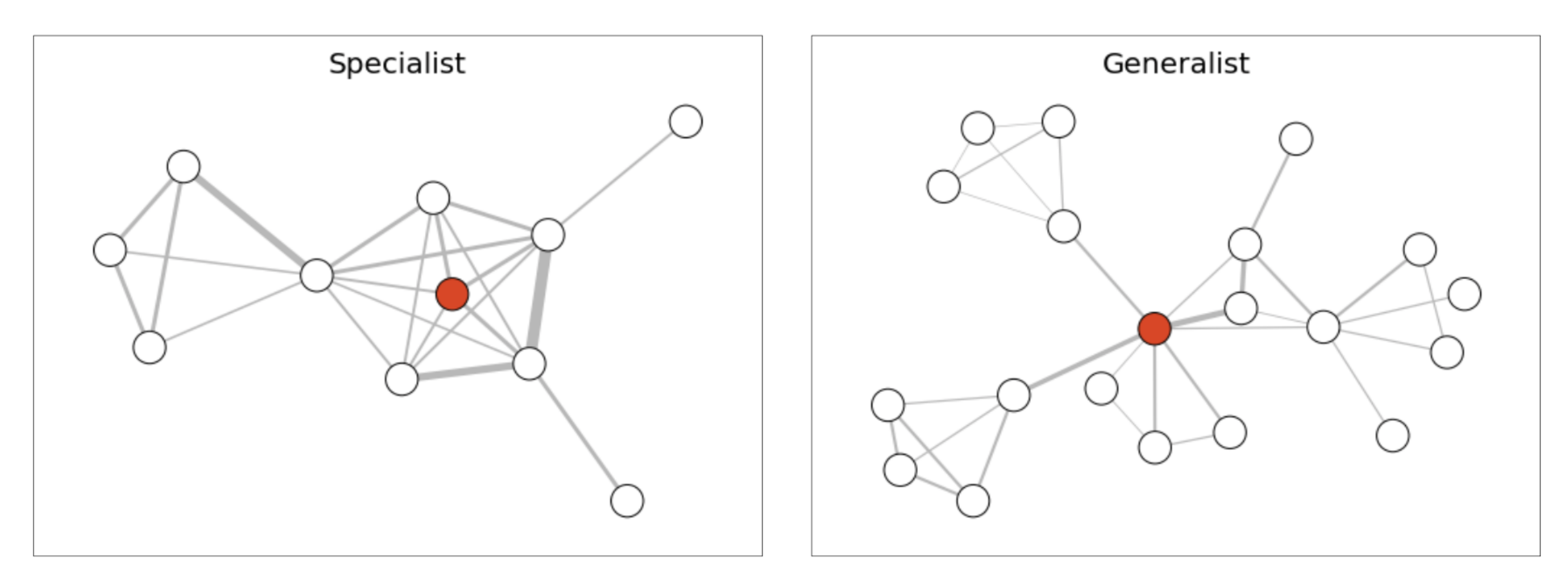}
    \caption{\textbf{Characteristic collaboration networks of specialists and generalists.} The two-step ego networks of a specialist and generalist criminal (red nodes) highlight characteristic differences in their collaboration networks. Generalists have larger, more open networks, while specialists have smaller, more closed networks characterized by repeated collaborations.}
    \label{fig:example_nets}
\end{figure}

Figure \ref{fig:example_nets} shows two characteristic ego networks for  specialists and generalists, providing a visual representation of the stylized patterns observed in Table \ref{tab:summstat}. The ego node is highlighted in red; we include all alters up to two steps away, as well as the connections between them. For instance, the specialist node is embedded in a clique: all five of their direct connections are themselves connected with each other. The thicker edges, apparent in the specialist's extended network, highlight repeated collaborations. The generalist's network, on the other hand, has significantly lower clustering. While the generalist has a higher degree (7 direct connections), there are fewer repeated connections and hardly any interactions among his direct neighbors themselves.

\begin{table}
\begin{tabular}{lrrrrrrr}
\toprule
\textbf{Specialists} &  Degree &  Strength &  2-Step Neighbors &  Clustering Coeff. &  Has Network Connection &  Strength/Degree \\
\midrule
mean  &         2.11 &           5.49 &             5.27 &                0.27 &             0.58 &          1.57 \\
std   &         5.41 &          16.04 &            11.84 &                0.42 &             0.49 &          5.44 \\
min   &         0 &           0 &             1 &                0 &             0 &          0 \\
25\%   &         0 &           0 &             1 &                0 &             0 &          0 \\
50\%   &         1 &           1 &             2 &                0 &             1 &          0.50 \\
75\%   &         2 &           6 &             5 &                0.60 &             1 &          1.75 \\
max   &       162 &         731 &           222 &                1 &             1 &        352 \\
\midrule
\textbf{Generalists} &  Degree &  Strength &  2-Step Neighbors &  Clustering Coeff. &  Has Network Connection &  Strength/Degree \\
\midrule
mean  &         2.98 &           3.26 &            11.74 &                0.23 &             0.66 &          0.66 \\
std   &         5.08 &           9.67 &            23.35 &                0.35 &             0.47 &          1.85 \\
min   &         0 &           0 &             1 &                0 &             0 &          0 \\
25\%   &         0 &           0 &             1 &                0 &             0 &          0 \\
50\%   &         1 &           1 &             3 &                0 &             1 &          0.50 \\
75\%   &         4 &           3 &            11 &                0.36 &             1 &          0.70 \\
max   &       117 &         581 &           404 &                1 &             1 &        118.25 \\

\bottomrule
\end{tabular}
\caption{\textbf{Specialists and generalists network position summary statistics.}}
\label{tab:summstat}
\end{table}

\section*{Discussion}

In this work we presented a method to define specialization in large datasets of criminal behavior. Motivated by the observation that defining specialization using legal code sections leads to artificial groupings we developed a method to cluster frequently co-occurring crimes types. The resulting clustering provides a useful data-defined categorization of crimes. It can also be used to define specialists in the population of criminals as those who stay within one category throughout their careers. We provide a new and effective method to cluster crime types and define specialization, that allows to identify specific socio-demographic and mobility characteristics of specialists and generalists. We further can locate specialists and generalist within their criminal collaboration network and interpret these positions.  

Our method to cluster crime types adapts and extends a statistical method from Tumminello et al. \cite{tumminello2013phenomenology}. Appropriately adopted to our data, we recover 21 crime clusters. Using the simple information theoretic concept if mutual information, we show that transitions are less likely out of certain crime types, suggesting that specialization is much more likely in certain categories than others. A strength of the new method is that each cluster can consist of a different number of crime types and number of offenses. Some crimes (such as fraud or drug possession) are much more common than others (counterfeiting or misuse of data).

Our main contribution is the presentation of large scale evidence on socio-demographic and behavioral differences between specialist and generalist individuals. We presented differences by gender and age, and indicated how specialists have different geographic ranges. The correlations between socio-demographic background and specialization suggest how background and socialization shape criminal behavior. Indeed, previous work on organized crime suggests that specialization happens over time (i.e. specialists will tend to be older) \cite{piquero1999onset}, that women have highly specialized roles \cite{tumminello2013phenomenology}, in line with our findings.

On the other hand, previous work on organized criminal groups suggests that localized activity tends to be generalist \cite{tumminello2021anagraphical}. In the case of our dataset that covers all crimes and not just organized crime as in \cite{tumminello2021anagraphical}, we find evidence that it is rather the specialists that tend to stay in the same place. Indeed, controlling for the number of crimes committed, we find that specialists have an 18\% lower radius of gyration than comparable generalists. That specialists tend to concentrate their activity in a specific region suggests that they rely on knowledge of a place and perhaps the support of individuals in a specific area to be effective. Criminologists have long understood criminal mobility in terms of opportunity: travel to a new place is costly and full of uncertainty \cite{townsley2016offender, mondani2022uncovering}. Our findings suggest that these costs are higher for specialized criminals. In other words, specialized criminal behavior may benefit from knowledge about a specific place or from repeated collaboration which is more easily coordinated in a small geographic area. One potential extension of our work would be to relate observed criminal mobility to the clustering of crimes in space \cite{prieto2018measuring}.

A further contribution is our finding that specialists and generalists have different collaboration patterns, measured via the position in their collaboration networks. Specialists have smaller but denser, more tightly knit collaboration networks. They are more likely to collaborate repeatedly with the same partners. This suggests that specialists are more effective when collaborating closely with others. For example: a high-level drug dealer may rely on others to launder profits. Previous work on specialization and collaboration among criminals has focused largely on organized crime as such organizations often exhibit a hierarchy of authority and a division of labor based on specialization and roles \cite{finckenauer2005problems}. For instance, specialization within a group is likely a signal of more sophisticated mafia activity \cite{tumminello2021anagraphical}. Criminals in the mafia are known to follow distinctive career paths shaped by their interactions with colleagues \cite{campedelli2021life}. Indeed, specialization (and the skills developed by specialists) and collaboration play complementary roles functioning of criminal networks \cite{duijn2014relative,sparrow1991application}. Our results suggest that specialists occupy important positions in general criminal networks, not only within specific organizations. Targeting specialists may prove an effective interdiction strategy as they are likely difficult to replace. Moreover, clusters of heterogeneous specialists working together may be highly suggestive of organized criminal activity. 

Our study has several limitations. As with nearly all empirical studies of criminal behavior, our data likely suffers from selection bias. In other words, our data does not contain information about undetected or unsolved cases of criminality. This is a common limitation in data-driven studies of criminal behavior, which by nature are limited to prosecuted or highly visible activities \cite{diviak2022tainted}. At the same time, events in our data are when individuals are charged with crimes - not all events are correctly assigned to the guilty individual. Second, although we have several years of activity, criminal careers can span decades. Data covering longer time periods could lead to deeper insights into individual paths through the world of crime. Finally, our data is not linked to information on incarceration. Certainly a longer period of time in jail or prison would limit an individual's ability to re-offend within our dataset. Future work should consider the consequences of incarceration itself on specialization, especially as an opportunity for transfer of human capital or strengthening of social capital between criminals \cite{damm2020prison}. Despite these limitations, our work provides new insights into socio-demographic, mobility, and collaborative characteristics of specialization in criminal careers at the scale of a whole country.

Our findings on specialization could also be extended to study organized crime. Indeed, standard economic theory suggests that specialization is a sign of division of labor \cite{smith2010wealth,becker1992division}. Criminals who focus exclusively on one kind of activity can become more efficient, but have to rely on others. For instance, a drug gang needs manufacturers, dealers, enforcers, and money launderers \cite{duijn2014relative}. Such inter-dependence requires trust, that depends on the nature of the collaboration \cite{von2004organized}. Some specialists will be more difficult to replace than others; these are natural targets for police intervention. Generalists, on the other hand, may be required to coordinate between distinct parts of criminal networks; these may be optimal targets to disintegrate networks into smaller pieces.





\bibliography{bibliography}

\section*{Acknowledgements}

We thank our project partner for access to data, feedback, and support.

\section*{Author contributions statement}

J.K., J.W. and S.T. conceived the experiments, G.H., J.K., T.P. and J.W. carried out the analyses. G.H., J.K., J.W. and S.T. wrote the manuscript. All authors reviewed the manuscript. 

\section*{Additional information}
\textbf{Competing interests}
The author(s) declare no competing interests.

\end{document}